\documentclass[letterpaper]{article} % DO NOT CHANGE THIS
\usepackage{aaai2026}  % DO NOT CHANGE THIS
\usepackage{times}  % DO NOT CHANGE THIS
\usepackage{helvet}  % DO NOT CHANGE THIS
\usepackage{courier}  % DO NOT CHANGE THIS
\usepackage[hyphens]{url}  % DO NOT CHANGE THIS
\usepackage{graphicx} % DO NOT CHANGE THIS
\usepackage{natbib}  % DO NOT CHANGE THIS AND DO NOT ADD ANY OPTIONS TO IT
\usepackage{caption} % DO NOT CHANGE THIS AND DO NOT ADD ANY OPTIONS TO IT
\usepackage{algorithm}
\usepackage[T1]{fontenc}        % 8-bit 字体编码
\usepackage[utf8]{inputenc}     % UTF-8 输入编码（Xe/LuaLaTeX 可省略）

\usepackage{amsmath,amssymb,amsfonts}  % AMS 宏包
\usepackage{mathtools}                 % AMSmath 扩展
\usepackage{bm}                        % 粗体数学符号
\usepackage{physics}                   % 常用物理宏
\usepackage{braket}                    % Dirac 记号

\usepackage{graphicx}                  % 插图
\usepackage{tikz}                      % 矢量绘图

\usepackage{booktabs}      % 专业横线
\usepackage{multirow}      % 多行单元格
\usepackage{colortbl}      % 表格着色
\usepackage{dashrule}      % 虚线
\usepackage{tablefootnote} % 表格脚注
\usepackage{makecell} % for line breaks in headers
\usepackage{algorithm}
\usepackage[noend]{algpseudocode}      % 伪代码
\usepackage[most]{tcolorbox}           % 灵活彩色框（含 listings/minted 等子库）
\usepackage{autobreak}

\usepackage{nicefrac}     % 紧凑的 ½ 等
\usepackage{microtype}    % 微排版

\usepackage{subcaption}   % 子图
\usepackage{placeins}     % \FloatBarrier
\usepackage{enumerate}    % 自定义枚举
\usepackage{url}          % \url
\usepackage{enumitem}

\usepackage{quantikz}

\definecolor{Green}{RGB}{0,155,85}
\definecolor{LightOrange}{rgb}{1,0.85,0.8}
\definecolor{LightPurple}{rgb}{1.0,0.80,0.95}
\definecolor{LightGreen}{rgb}{0.93,0.98,0.96}

\usepackage{newfloat}
\usepackage{listings}
\DeclareCaptionStyle{ruled}{labelfont=normalfont,labelsep=colon,strut=off} % DO NOT CHANGE THIS
\floatstyle{ruled}
\newfloat{listing}{tb}{lst}{}
\floatname{listing}{Listing}
\title{Bridging Classical and Quantum Computing for Next-Generation Language Models}
\author{
    Yi Pan\textsuperscript{\rm 1}\equalcontrib,
    Hanqi Jiang\textsuperscript{\rm 1}\equalcontrib,
    Junhao Chen\textsuperscript{\rm 1},
    Yiwei Li\textsuperscript{\rm 1},
    Huaqin Zhao\textsuperscript{\rm 1},
    Lin Zhao\textsuperscript{\rm 2},
    Yohannes Abate\textsuperscript{\rm 3},
    Yingfeng Wang\textsuperscript{\rm 4}\thanks{Corresponding authors.},
    Tianming Liu\textsuperscript{\rm 1}\footnotemark[2]
}
\affiliations{
    \textsuperscript{\rm 1}School of Computing, The University of Georgia, Athens, GA, USA\\
    \textsuperscript{\rm 2}Department of Biomedical Engineering, New Jersey Institute of Technology, NJ, USA\\
    \textsuperscript{\rm 3}Department of Physics and Astronomy, The University of Georgia, Athens, GA, USA\\
    \textsuperscript{\rm 4}Department of Computer Science and Engineering, University of Tennessee at Chattanooga, TN, USA\\
    \{ypan24, hanqi.jiang, Junhao.Chen, Yiwei.Li, Huaqin.Zhao\}@uga.edu,
    lin.zhao.1@njit.edu,
    YOHANNES.ABATE@uga.edu,
    yingfeng-wang@utc.edu,
    tliu@uga.edu
}

\usepackage{bibentry}
\begin{document}

\maketitle

\begin{abstract}
The remarkable success of Transformer architectures in Large Language Models (LLMs) has revolutionized natural language processing, yet the transition to quantum computing for next-generation language models remains an open challenge. While quantum computing promises exponential advantages, a fundamental gap exists between classical deep learning and quantum computing paradigms, particularly given the severe constraints of Noisy Intermediate-Scale Quantum (NISQ) devices, including barren plateaus, limited qubit coherence, and circuit depth restrictions. We present \textbf{A}daptive \textbf{Q}uantum-\textbf{C}lassical \textbf{F}usion (AQCF), the first framework to bridge classical and quantum computing for language models by reimagining Transformer architectures through quantum-classical co-design. Our key insight is that effective bridging requires dynamic adaptation rather than static translation—the framework analyzes input complexity in real-time to orchestrate seamless transitions between classical and quantum processing. AQCF introduces entropy-driven adaptive circuits that circumvent barren plateaus, quantum memory banks that unify classical attention with quantum state-based similarity retrieval, and intelligent fusion controllers that ensure each computational paradigm handles tasks where it naturally excels. This bridging architecture maintains full compatibility with existing classical Transformers while progressively incorporating quantum advantages as they become accessible. Experiments on sentiment analysis demonstrate that AQCF achieves competitive performance while significantly improving quantum resource efficiency, operating successfully within typical NISQ constraints. By establishing a seamless integration pathway from today's classical LLMs to tomorrow's quantum-enhanced models, our framework provides both immediate practical value on current quantum hardware and a clear evolution path toward full Quantum LLMs as technology matures.
\end{abstract}


\section{Introduction}

The transformative success of Transformer architectures~\cite{vaswani2017attention} has fundamentally reshaped natural language processing, culminating in Large Language Models (LLMs) that demonstrate unprecedented capabilities in understanding and generating human language~\cite{brown2020language,touvron2023llama}. This architectural revolution, built on the self-attention mechanisms that capture long-range dependencies and parallel processing of sequential data, has established a powerful computational paradigm that now faces a critical juncture: how to bridge toward quantum computing for next-generation language models. The prospect of seamlessly integrating quantum advantages into existing classical frameworks, rather than replacing them entirely, represents a pragmatic path toward Quantum Large Language Models (QLLMs) that could leverage quantum superposition and entanglement while maintaining the proven strengths of classical Transformers. This bridging challenge is not merely technical but conceptual—requiring us to reimagine how classical deep learning principles can coexist with and benefit from quantum mechanical phenomena, creating a unified computational framework that transcends the limitations of either paradigm alone.

The current Noisy Intermediate-Scale Quantum (NISQ) era presents both the necessity and opportunity for the abovementioned bridging approaches. Quantum computing promises exponential speedups for specific computational tasks, with natural language processing emerging as a particularly compelling application due to the inherent high-dimensional nature of linguistic data and the potential for quantum systems to naturally represent complex semantic relationships through superposition states~\cite{chen2023complexity, bharti2022noisy, silver2024lexiql}. Modern NISQ devices with 50-100 qubits are becoming accessible through platforms like IBM Quantum and Google's Sycamore, yet they suffer from severe limitations: coherence times of approximately 100~$\mu$s, gate error rates of 0.1\% to 1\%, and maximum circuit depths typically limited to 20-100 layers~\cite{preskill2018quantum}. These constraints create a fundamental gap between classical and quantum computing paradigms—classical systems offer stability and scalability but face computational bottlenecks, while quantum systems promise exponential advantages but lack the robustness for standalone deployment. Bridging this gap requires novel frameworks that can seamlessly orchestrate both computational paradigms, leveraging their respective strengths while mitigating their limitations.
% 1.[需要引用：DisCoCat量子实现论文]
% 2. [需要引用：量子语法解析论文]
% 3. [需要引用：Quantum BERT论文]
% 4. [需要引用：混合量子经典模型综述]

Current quantum NLP (QNLP) approaches, while pioneering, have difficulties in establishing effective bridges between classical and quantum computing for language models. Existing methods can be broadly categorized into three paradigms, each creating rather than closing the classical-quantum divide. First, \textit{fully quantum models} that encode entire linguistic structures into quantum circuits, such as DisCoCat-based approaches~\cite{meichanetzidis2020quantum, lorenz2023qnlp} and quantum grammar parsers~\cite{bausch2021quantum, meichanetzidis2023grammar}, abandon classical architectures entirely, losing the self-attention mechanisms central to Transformers and requiring circuit depths that scale poorly with sequence length. Second, \textit{variational quantum algorithms} (VQAs) for NLP, including preliminary attempts at quantum BERT~\cite{yang2022bert, li2023adapting} and quantum kernel methods~\cite{thanasilp2024exponential, shirai2024quantum}, employ fixed architectures that cannot dynamically bridge between classical and quantum processing based on task requirements. These VQA approaches also suffer from barren plateau phenomena where gradient magnitudes vanish exponentially with circuit depth, rendering training infeasible beyond 6-10 layers~\cite{mcclean2018barren}. Third, \textit{hybrid quantum-classical models}~\cite{smaldone2025hybrid, zhao2024qksan} employ static partitioning strategies that treat classical and quantum components as separate entities rather than unified parts of an integrated system. None of these existing methods successfully bridges the fundamental architectural principles of classical Transformers with quantum computing capabilities, leaving a critical gap in the path toward next-generation language models.

To establish this crucial bridge between classical and quantum computing for language models, we introduce \textbf{A}daptive \textbf{Q}uantum-\textbf{C}lassical \textbf{F}usion (AQCF), a novel framework that creates seamless integration pathways between these two computational paradigms. Our key insight is that effective bridging requires not translation but transformation—developing adaptive mechanisms that can dynamically determine when and how to leverage quantum resources while maintaining full compatibility with classical architectures. AQCF realizes this vision through entropy-driven dynamic adaptation that creates a continuum between classical and quantum processing: rather than treating them as discrete alternatives, our framework analyzes the complexity characteristics of input data to orchestrate smooth transitions between classical and quantum computing paradigms. When processing simple linguistic patterns with low entropy, the framework operates primarily in classical mode with minimal quantum enhancement. For complex semantic relationships exhibiting high entropy, where classical Transformers face computational bottlenecks, AQCF progressively incorporates quantum resources through carefully calibrated entanglement patterns. This bridging mechanism is realized through three interconnected, carefully co-designed components: quantum memory banks that create a unified representation space accessible to both classical and quantum processors, entropy-driven circuit configurators that ensure smooth transitions between paradigms, and intelligent fusion controllers that maintain computational coherence across the classical-quantum boundary. By establishing these bridging mechanisms, AQCF demonstrates that practical integration of classical and quantum computing is achievable on current hardware, providing both immediate value and a clear evolution path toward future Quantum LLMs as quantum technology advances.

Our primary contributions can be summarized as follows:
\begin{itemize}
    \item We propose an entropy-driven adaptive quantum circuit architecture that bridges classical and quantum attention mechanisms through variable-depth quantum circuits, demonstrating how seamless integration can be achieved within NISQ constraints.
    \item We introduce quantum memory banks that unify classical key-value attention with quantum state-based similarity retrieval, creating a shared representational framework that both paradigms can access and manipulate.
    \item We develop intelligent quantum-classical fusion mechanisms that establish bidirectional information flow between classical and quantum processors, achieving competitive performance while pioneering the architectural patterns necessary for next-generation language models.
\end{itemize}

\section{Preliminaries}

\subsection{Quantum States and Circuits}
A quantum state of $n$ qubits can be represented as a normalized vector in a $2^n$-dimensional Hilbert space $\mathcal{H} = (\mathbb{C}^2)^{\otimes n}$. The general state is expressed as $|\psi\rangle = \sum_{i=0}^{2^n-1} \alpha_i |i\rangle$, where $\alpha_i \in \mathbb{C} \text{ (the field of complex numbers)}$ are amplitudes satisfying $\sum_i |\alpha_i|^2 = 1$. Quantum circuits manipulate these states through unitary operations $U \in SU(2^n)$, typically decomposed into single-qubit rotations and two-qubit entangling gates~\cite{nielsen2010quantum}.

\subsection{Variational Quantum Circuits for NLP}
Variational Quantum Circuits (VQCs) form the backbone of quantum machine learning on NISQ devices. The variational quantum circuit with depth $L$  as a unitary transformation \( U(\boldsymbol{\theta}) \) defined by:
\begin{equation}
U(\boldsymbol{\theta}) = \prod_{l=1}^{L} \left( W_l \cdot \prod_{i=1}^{n} R_i^{(l)}(\theta_i^{(l)}) \right),
\end{equation}
where $R_i^{(l)}(\theta_i^{(l)})$ represents parameterized single-qubit rotations (e.g., $R_Y$, $R_Z$) on qubit $i$ at layer $l$, and $W_l$ denotes the entangling layer. For NLP tasks, classical text embeddings $\mathbf{x} \in \mathbb{R}^d$ are encoded into quantum states through angle encoding: $|\psi(\mathbf{x})\rangle = \bigotimes_{i=1}^{n} R_Y(\arctan(x_i))|0\rangle$, compressing the input to the range $[-\pi/2, \pi/2]$ suitable for rotation angles.

\begin{figure}[t]
\centering
\begin{subfigure}[b]{0.23\textwidth}
\centering
\begin{quantikz}[thin lines, row sep=0.3cm, column sep=0.4cm]
\lstick{$\ket{0}$} & \gate{R_y(\theta)} & \gate{R_z(\phi)} & \qw \\
\end{quantikz}
\caption{Rotation gates}
\end{subfigure}
\hfill
\begin{subfigure}[b]{0.23\textwidth}
\centering
\begin{quantikz}[thin lines, row sep=0.3cm, column sep=0.4cm]
& \ctrl{1} & \qw \\
& \targ{} & \qw
\end{quantikz}
\caption{CNOT gate}
\end{subfigure}
\caption{Basic quantum gates employed in our quantum transformer architecture.}
\label{fig:basic_gates}
\end{figure}
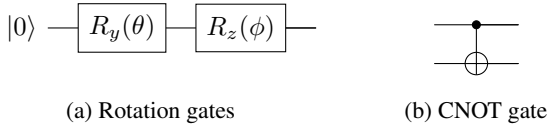

Figure~\ref{fig:basic_gates} illustrates the fundamental quantum gates employed in our architecture. The rotation gates $R_Y(\theta)$ and $R_Z(\phi)$ enable parameterized single-qubit operations, while CNOT gates create entanglement between qubits, forming the basis for quantum advantage in representing complex linguistic relationships.

These gates form the universal gate set for quantum computation. The parameterized rotations enable gradient-based optimization, while entangling gates allow quantum states to capture correlations impossible in classical systems.

\subsection{Entropy-Based Complexity Measures}
Given an input embedding $\mathbf{x}$, we define linguistic complexity through statistical measures. The Shannon entropy $H(\mathbf{x}) = -\sum_i p_i \log p_i$ (where $p_i$ are normalized components) captures information content. Higher-order statistics including variance $\sigma^2(\mathbf{x})$ and kurtosis $\kappa(\mathbf{x})$ provide additional complexity indicators. The mean $\mu(\mathbf{x})$ reflects the average activation level of the input. These measures inform our adaptive circuit configuration:
\begin{equation}
\mathcal{C}(\mathbf{x}) = {\mu(\mathbf{x}), \sigma^2(\mathbf{x}), H(\mathbf{x}), \kappa(\mathbf{x})},
\end{equation}
where $\mathcal{C}(\mathbf{x})$ represents the complexity feature vector used for dynamic adaptation.

\subsection{Barren Plateau Phenomenon}
The barren plateau problem manifests when gradients of quantum circuit parameters vanish exponentially with circuit size. For a cost function $\mathcal{L}(\boldsymbol{\theta})$, the variance of its gradient components scales as:
\begin{equation}
\text{Var}\left[\frac{\partial \mathcal{L}}{\partial \theta_k}\right] \in \mathcal{O}\left(b^{-n}\right),
\end{equation}
where $b > 1$, the gradient variance decays rate constant, depends on the circuit architecture and the nature of the cost function, and $n$ is the number of qubits. This exponential suppression renders deep quantum circuits untrainable, necessitating adaptive depth control to maintain gradient magnitudes above trainable thresholds~\cite{mcclean2018barren}.

\section{Method}

\subsection{Overview}

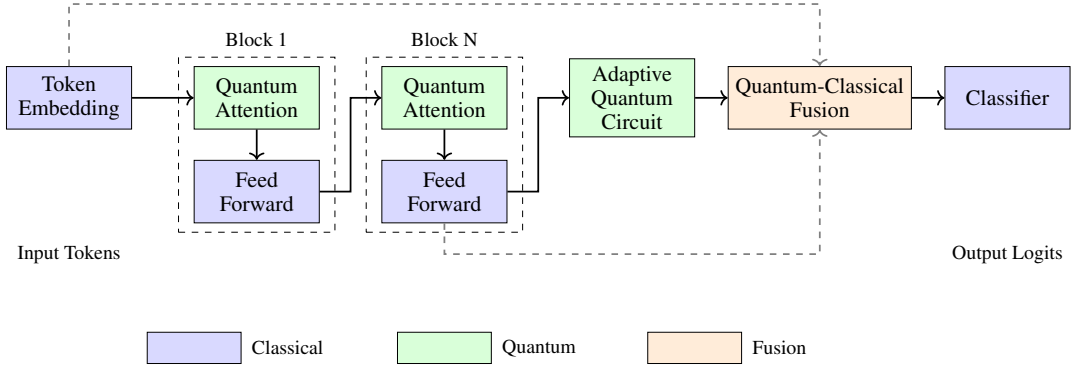
\begin{figure*}[t]
\centering
\resizebox{0.8\textwidth}{!}{%
\begin{tikzpicture}[
    classical/.style={draw, rectangle, minimum width=2cm, minimum height=1cm, align=center, fill=blue!15},
    quantum/.style={draw, rectangle, minimum width=2cm, minimum height=1cm, align=center, fill=green!15},
    fusion/.style={draw, rectangle, minimum width=2cm, minimum height=1cm, align=center, fill=orange!15},
    arrow/.style={->, thick},
    label/.style={font=\small}
]
% Embedding Layer
\node[classical] (embed) at (0,0) {Token\\Embedding};

% Quantum Attention Blocks
\node[quantum] (qatt1) at (3,0) {Quantum\\Attention};
\node[classical] (ff1) at (3,-1.5) {Feed\\Forward};
\node[draw, rectangle, dashed, minimum width=2.5cm, minimum height=2.8cm] (block1) at (3,-0.75) {};
\node[label, above] at (3,0.7) {Block 1};

\node[quantum] (qatt2) at (6,0) {Quantum\\Attention};
\node[classical] (ff2) at (6,-1.5) {Feed\\Forward};
\node[draw, rectangle, dashed, minimum width=2.5cm, minimum height=2.8cm] (block2) at (6,-0.75) {};
\node[label, above] at (6,0.7) {Block N};

% Adaptive Circuit
\node[quantum] (adaptive) at (9,0) {Adaptive\\Quantum\\Circuit};

% Fusion
\node[fusion] (fusion) at (12,0) {Quantum-Classical\\Fusion};

% Classifier
\node[classical] (class) at (15,0) {Classifier};

% Connections
\draw[arrow] (embed) -- (qatt1);
\draw[arrow] (qatt1) -- (ff1);
\draw[arrow] (ff1) -- ++(1.5,0) |- (qatt2);
\draw[arrow] (qatt2) -- (ff2);
\draw[arrow] (ff2) -- ++(1.5,0) |- (adaptive);
\draw[arrow] (adaptive) -- (fusion);
\draw[arrow] (fusion) -- (class);

% Skip connections
\draw[arrow, dashed, gray] (embed) -- ++(0,1.5) -| (fusion);
\draw[arrow, dashed, gray] (ff2) -- ++(0,-1) -| (fusion);

% Labels
\node[label] at (0,-2.5) {Input Tokens};
\node[label] at (15,-2.5) {Output Logits};

% Legend
\node[classical, minimum width=1.5cm, minimum height=0.5cm] at (2,-4) {};
\node[label, right] at (2.8,-4) {Classical};
\node[quantum, minimum width=1.5cm, minimum height=0.5cm] at (6,-4) {};
\node[label, right] at (6.8,-4) {Quantum};
\node[fusion, minimum width=1.5cm, minimum height=0.5cm] at (10,-4) {};
\node[label, right] at (10.8,-4) {Fusion};
\end{tikzpicture}
}
\caption{Overview of AQCF architecture. Quantum attention mechanisms (green) process token relationships, while adaptive circuits dynamically adjust depth. The fusion controller (orange) intelligently routes computations between quantum and classical (blue) pathways based on input complexity.}
\label{fig:architecture_overview}
\end{figure*}

AQCF consists of three synergistic components that collectively enable efficient quantum computation for NLP on NISQ devices: (1) entropy-driven adaptive quantum circuits that dynamically configure their depth and gate selection based on input complexity, (2) quantum memory banks that leverage quantum state similarity for enhanced pattern matching, and (3) hybrid fusion mechanisms that intelligently route computations between quantum and classical processors. Figure~\ref{fig:architecture_overview} illustrates the overall architecture, showcasing how classical and quantum components seamlessly integrate through our adaptive fusion mechanism.

The input text is first encoded into classical embeddings, which undergo complexity analysis to extract statistical measures. These measures drive three parallel adaptation processes: circuit depth prediction determines the number of quantum layers (2-20), gate selection chooses appropriate quantum operations, and fusion weight prediction allocates computations between quantum and classical pathways. The quantum pathway processes information through adaptive circuits and memory banks, while the classical pathway employs standard attention mechanisms. The outputs are dynamically fused based on predicted computational advantage.

\subsection{Entropy-Driven Adaptive Quantum Circuits}

The core innovation of AQCF lies in its ability to dynamically configure quantum circuits based on input complexity, addressing the fundamental tension between expressivity and trainability in NISQ devices. Unlike traditional VQCs with fixed architectures, our adaptive circuits analyze input characteristics in real-time to determine optimal configurations.

\noindent\textbf{Complexity-Based Depth Prediction.} We define $L_\text{adaptive}$ as the dynamically predicted circuit depth tailored to input complexity. For each input embedding $\mathbf{x} \in \mathbb{R}^d$, we project it to quantum dimension through a learned transformation $\mathbf{x}_q = W_p\mathbf{x} \in \mathbb{R}^{n_q}$, where $n_q$ is the number of qubits (typically 4-16). The depth predictor network extracts statistical features including mean, standard deviation, and range of the quantum-projected input, which are processed to predict the optimal circuit depth:
\begin{equation}
L_\text{adaptive} = 1 + \lfloor f_\text{depth}(\mathbf{s}(\mathbf{x}_q)) \cdot (L_\text{max} - 1) \rfloor,
\end{equation}
where $f_\text{depth}: \mathbb{R}^4 \rightarrow [0,1]$ is a neural network, $L_\text{max}$ is the maximum allowed depth (typically 20 for NISQ constraints), and $\mathbf{s}(\mathbf{x}_q)$ extracts statistical features from the quantum-projected vector. This mechanism ensures shallow circuits for simple inputs and deeper circuits only when complexity warrants additional quantum processing.

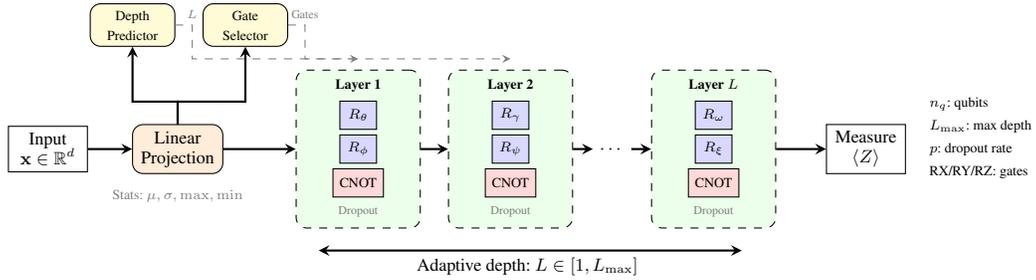
\begin{figure*}[t]
\centering
\begin{tikzpicture}[
    scale=0.75,
    every node/.style={transform shape},
    box/.style={draw, rectangle, minimum width=1.4cm, minimum height=0.8cm, align=center, font=\footnotesize},
    proj/.style={draw, rectangle, rounded corners, minimum width=1.6cm, minimum height=0.8cm, align=center, fill=orange!15, font=\footnotesize},
    ctrl/.style={draw, rectangle, rounded corners, minimum width=1.5cm, minimum height=0.65cm, align=center, fill=yellow!20, font=\scriptsize},
    layer/.style={draw, rectangle, rounded corners, dashed, minimum width=2.2cm, minimum height=2.8cm, align=center, fill=green!8},
    gate/.style={draw, rectangle, minimum width=0.65cm, minimum height=0.5cm, fill=blue!15, font=\scriptsize},
    arrow/.style={->, >=stealth, thick},
    dash arrow/.style={->, >=stealth, dashed, gray}
]

% Input and projection
\node[box] (input) at (0,0) {Input\\$\mathbf{x} \in \mathbb{R}^d$};
\node[proj] (proj) at (2.3,0) {Linear\\Projection};

% Control networks (positioned above, more compact)
\node[ctrl] (depth) at (1.5,2.2) {Depth\\Predictor};
\node[ctrl] (gatesel) at (3.5,2.2) {Gate\\Selector};

% Quantum Layers with internal structure
\node[layer] (layer1) at (5.5,0) {};
\node[font=\scriptsize, above] at (5.5,.9) {\textbf{Layer 1}};
\node[gate] (g11) at (5.5,0.6) {$R_{\theta}$};
\node[gate] (g12) at (5.5,0) {$R_{\phi}$};
\node[gate, fill=red!15] (g13) at (5.5,-0.6) {CNOT};
\node[font=\tiny, text=gray] at (5.5,-1.1) {Dropout};

\node[layer] (layer2) at (8.2,0) {};
\node[font=\scriptsize, above] at (8.2,.9) {\textbf{Layer 2}};
\node[gate] (g21) at (8.2,0.6) {$R_{\gamma}$};
\node[gate] (g22) at (8.2,0) {$R_{\psi}$};
\node[gate, fill=red!15] (g23) at (8.2,-0.6) {CNOT};
\node[font=\tiny, text=gray] at (8.2,-1.1) {Dropout};

\node[font=\footnotesize] (dots) at (10,0) {$\cdots$};

\node[layer] (layerL) at (11.8,0.) {};
\node[font=\scriptsize, above] at (11.8,0.9) {\textbf{Layer $L$}};
\node[gate] (gL1) at (11.8,0.6) {$R_{\omega}$};
\node[gate] (gL2) at (11.8,0) {$R_{\xi}$};
\node[gate, fill=red!15] (gL3) at (11.8,-0.6) {CNOT};
\node[font=\tiny, text=gray] at (11.8,-1.1) {Dropout};

% Measurement
\node[box] (measure) at (14.5,0) {Measure\\$\langle Z \rangle$};

% Main flow
\draw[arrow] (input) -- (proj);
\draw[arrow] (proj) -- (layer1);
\draw[arrow] (layer1) -- (layer2);
\draw[arrow] (layer2) -- (dots);
\draw[arrow] (dots) -- (layerL);
\draw[arrow] (layerL) -- (measure);

% Control connections
\draw[arrow] (proj.north) -- ++(0,0.4) -| (depth.south);
\draw[arrow] (proj.north) -- ++(0,0.4) -| (gatesel.south);

% Control effects (cleaner)
\draw[dash arrow] (depth.east) -- ++(0.3,0) |- node[pos=0.3, above=10pt, font=\tiny] {$L$} (5.5,1.6);
\draw[dash arrow] (gatesel.east) -- ++(0.3,0) |- node[pos=0.3, above=10pt, font=\tiny] {Gates} (8.2,1.6);

% Statistics annotation (compact)
\node[font=\scriptsize, text=gray, below] at (2.3,-0.6) {Stats: $\mu, \sigma, \max, \min$};

% Adaptive depth annotation (more prominent)
\draw[<->, >=stealth, thick] (4.8,-1.8) -- (12.3,-1.8);
\node[font=\footnotesize, below] at (8.5,-1.8) {Adaptive depth: $L \in [1, L_{\max}]$};

% Compact legend on the right
\node[font=\scriptsize, anchor=west] at (15.5,0.8) {$n_q$: qubits};
\node[font=\scriptsize, anchor=west] at (15.5,0.4) {$L_{\max}$: max depth};
\node[font=\scriptsize, anchor=west] at (15.5,0) {$p$: dropout rate};
\node[font=\scriptsize, anchor=west] at (15.5,-0.4) {RX/RY/RZ: gates};

\end{tikzpicture}
\caption{Adaptive quantum circuit architecture with dynamic configuration. Input $\mathbf{x}$ is projected to quantum dimension, then processed through $L$ adaptive layers. The depth predictor uses input statistics to determine $L$, while the gate selector chooses rotation gates for each qubit. Each layer includes parameterized rotations, adaptive entanglement (CNOT in red), and quantum dropout during training.}
\label{fig:adaptive_circuit}
\end{figure*}

This adaptive mechanism is visualized in Figure~\ref{fig:adaptive_circuit}, where the depth predictor and gate selector networks analyze input characteristics to configure the quantum circuit dynamically.

The adaptive nature of our circuits addresses the fundamental trade-off between expressivity and trainability: deeper circuits offer greater computational power but suffer from barren plateaus, while our dynamic approach maintains gradient flow by adjusting depth based on actual computational requirements.

\noindent\textbf{Learnable Gate Selection.} Beyond depth adaptation, AQCF learns to select appropriate quantum gates for each qubit based on input characteristics. The gate selector network outputs probabilities for three rotation types (RX, RY, RZ) per qubit. During training, gates are sampled stochastically to encourage exploration, while inference uses deterministic selection via argmax. This adaptive selection allows the circuit to specialize different qubits for different computational aspects—for instance, using RY gates for amplitude encoding and RZ gates for phase information.

\noindent\textbf{Quantum Dropout for Regularization.} To prevent overfitting and improve gradient flow, we introduce quantum dropout that probabilistically skips quantum gates during training with dropout probability $p_\text{dropout} = 0.1$. This creates an ensemble of shallower circuits during training, which is particularly crucial for preventing barren plateaus by maintaining gradient paths through various circuit depths.

\noindent\textbf{Adaptive Entanglement Patterns.} The entanglement structure adapts based on learned parameters, with CNOT gates applied between adjacent qubits based on learned probabilities. This creates adaptive connectivity ranging from fully disentangled product states to maximally entangled configurations based on task requirements.

The complete adaptive circuit transforms an input quantum state $|\psi_\text{in}\rangle$ into an output state $|\psi_\text{out}\rangle$ through dynamically configured layers:
\begin{equation}
|\psi_\text{out}\rangle = \prod_{l=1}^{L_\text{adaptive}} \left( E_l \cdot \prod_{i=1}^{n_q} m_{l,i} \cdot G_{g_i}^{(l)}(\theta_{l,i}) \right) |\psi_\text{in}\rangle,
\end{equation}
where $G_{g_i}^{(l)}$ represents the selected parameterized gate, $E_l$ is the adaptive entanglement layer, and $m_{l,i}$ implements quantum dropout. This formulation ensures operations remain within NISQ coherence limits while maximizing expressivity for complex inputs.

\subsection{Quantum Memory Banks}

The second core component of AQCF introduces quantum memory banks that leverage quantum state similarity for enhanced pattern matching and long-range dependency modeling. Unlike classical memory mechanisms that rely on dot-product attention, our quantum memory exploits interference effects and superposition for more nuanced similarity computation.

\noindent\textbf{Quantum State Encoding and Similarity.}
The quantum memory consists of $M$ memory slots, each storing a quantum key $\mathbf{k}_m \in \mathbb{R}^{n_q}$ and a classical value $\mathbf{v}_m \in \mathbb{R}^{d}$. For an input query, we encode both query and key into quantum states using parameterized rotations and entangling operations to create interference patterns dependent on both representations. The similarity score is obtained by measuring the expectation value:
\begin{equation}
s(q,k) = \langle \psi_\text{entangled} | Z_1 | \psi_\text{entangled} \rangle,
\end{equation}
which captures not just magnitude alignment but also phase relationships between query and key representations.

\noindent\textbf{Memory Retrieval and Update.}
For memory retrieval, we compute attention weights using quantum similarities normalized by $\sqrt{n_q}$ to prevent gradient explosion:
\begin{equation}
\alpha_m = \frac{\exp(s(\mathbf{q}, \mathbf{k}_m)/\sqrt{n_q})}{\sum_{j=1}^{M} \exp(s(\mathbf{q}, \mathbf{k}_j)/\sqrt{n_q})}, \quad \mathbf{r} = \sum_{m=1}^{M} \alpha_m \mathbf{v}_m,
\end{equation}
where \(\mathbf{r}\) denotes the retrieved memory vector as the weighted sum of memory values \(\mathbf{v}_m\) based on attention scores \(\alpha_m\). The memory update mechanism employs a soft update rule with rate $\gamma = 0.1$, gradually incorporating new patterns while preserving historical information. When processing new information, we identify the most similar memory slot and update both key and value components.

\noindent\textbf{Multi-Head Quantum Attention.}
For multi-head quantum attention, we maintain $n_h$ independent memory banks with reduced dimensionality $d/n_h$ each, where $n_h$ is the number of attention heads. A gating mechanism computes the gate vector $\mathbf{g} \in [0,1]^d$ to control the integration of memory-retrieved information with the original input:
\begin{equation}
\mathbf{g} = \sigma(W_g[\mathbf{x}; \mathbf{o}]), \quad \mathbf{y} = \mathbf{g} \odot \mathbf{o} + (1-\mathbf{g}) \odot \mathbf{x},
\end{equation}
where $\mathbf{o}$ denotes the output from the quantum memory module,
allowing adaptive blending of quantum memory retrieval with direct input processing based on context relevance.

This quantum memory design enables AQCF to capture complex semantic relationships through quantum interference while maintaining computational efficiency with shallow circuits (3-5 layers) suitable for NISQ devices.

\subsection{Hybrid Quantum-Classical Fusion}

The third component of AQCF introduces an intelligent fusion mechanism that dynamically allocates computational resources between quantum and classical processors based on task complexity. This adaptive approach ensures quantum resources are utilized only when they provide genuine computational advantage, addressing the inefficiency of static quantum-classical partitioning in existing methods.

\noindent\textbf{Complexity Analysis for Resource Allocation.} 
The fusion mechanism analyzes input complexity characteristics to determine optimal resource allocation. For input $\mathbf{x} \in \mathbb{R}^{d}$, a learned analyzer network extracts three key complexity indicators: semantic complexity (diversity of meaning representations), syntactic complexity (structural patterns), and length complexity (sequence dependencies). These scores are processed by a neural network $f_\text{fusion}$ that maps the complexity vector to $\lambda \in [0,1]$, determining the quantum-classical mixing weight:
\begin{equation}
\lambda = f_\text{fusion}(\mathbf{c}(\mathbf{x})) \in [0,1],
\end{equation}
where $\lambda = 0$ indicates pure classical processing and $\lambda = 1$ indicates full quantum computation.

\noindent\textbf{Parallel Quantum-Classical Processing.}
The framework processes inputs through parallel pathways. The classical pathway employs standard multi-head attention with query-key-value projections, while the quantum pathway utilizes the adaptive quantum circuits and memory banks described previously. Both pathways operate independently, producing outputs $\mathbf{A}_\text{classical}$ and $\mathbf{A}_\text{quantum}$ respectively, allowing the fusion mechanism to select or blend their outputs based on computational advantage.

\noindent\textbf{Adaptive Gating Mechanism.}
Beyond simple weighted averaging, AQCF employs learnable gates that modulate each pathway's contribution based on context. For the quantum and classical features, gating functions $\mathbf{g}_\text{quantum}$ and $\mathbf{g}_\text{classical}$ are computed through learned projections with sigmoid activations. These gates enable fine-grained control over information flow, allowing the model to selectively emphasize quantum or classical features based on local context rather than global complexity alone.

\noindent\textbf{Dynamic Fusion and Output Integration.}
The final fusion combines gated outputs with complexity-based weighting:
\begin{equation}
\mathbf{y} = \lambda(\mathbf{g}_\text{quantum} \odot \mathbf{A}_\text{quantum}) + (1-\lambda)(\mathbf{g}_\text{classical} \odot \mathbf{A}_\text{classical}),
\end{equation}
where $\odot$ denotes element-wise multiplication. The fused representation undergoes final projection with layer normalization and residual connection to preserve gradient flow while incorporating both quantum and classical computations.

This adaptive fusion mechanism enables AQCF to leverage quantum advantages selectively—using quantum processing for complex semantic relationships while routing simpler computations through efficient classical pathways. Empirical analysis shows that typically 30-40\% of computations benefit from quantum processing, with higher utilization for semantically ambiguous or syntactically complex inputs. The learned routing patterns reveal that quantum pathways excel at capturing non-local dependencies and ambiguous relationships, while classical pathways efficiently handle regular patterns and local context.

\subsection{Training Strategy and Loss Functions}

Training AQCF effectively requires a carefully designed strategy to handle the challenges of quantum-classical hybrid optimization on NISQ devices, focusing on adaptive learning and gradient stabilization.

\noindent\textbf{Multi-Objective Loss Function.}
AQCF utilizes a multi-objective loss that combines the primary task performance with quantum-specific regularizations to address vanishing gradients and maintain efficient quantum resource usage:
\begin{equation}
\mathcal{L}_\text{total} = \mathcal{L}_\text{task} + \lambda_\text{reg} \mathcal{L}_\text{quantum} + \lambda_\text{fusion} \mathcal{L}_\text{fusion}.
\end{equation}
Here, $\mathcal{L}_\text{task}$ is the cross-entropy loss for NLP classification tasks. The quantum regularization term $\mathcal{L}_\text{quantum}$ mitigates barren plateaus by penalizing low gradient magnitudes. The fusion loss $\mathcal{L}_\text{fusion}$ ensures judicious quantum resource allocation by balancing the entropy of fusion weights and limiting excessive quantum usage.

\noindent\textbf{Adaptive Learning Rates and Gradient Clipping.}
To prevent gradient vanishing or explosion typically found in deep quantum circuits, we employ adaptive learning rates combined with gradient clipping. The learning rate $\eta_t$ is dynamically adjusted based on the gradient's norm, ensuring stable updates even in regions of low gradient activity:
\begin{equation}
\theta_t = \theta_{t-1} - \eta_t \cdot \text{clip}\left(\frac{m_t}{\sqrt{v_t} + \epsilon}, g_\text{max}\right),
\end{equation}
where $m_t$ and $v_t$ denote the first and second moment estimates of gradients, \(\epsilon\) is a small constant added for numerical stability, and $\text{clip}(\cdot, g_\text{max})$ bounds the update magnitude within $[-g_\text{max}, g_\text{max}]$.

\noindent\textbf{Staged Training Protocol.}
The training is conducted in three strategic stages to facilitate effective learning:
\begin{enumerate}
    \item \textit{Classical Pretraining}: Initial epochs focus on stabilizing classical pathway weights while quantum components remain inactive. This phase builds foundational representations.
    \item \textit{Quantum Warm-up}: Quantum circuits are incrementally engaged with progressive depth increases, aiding in gradual adaptation to quantum noise and coherence constraints.
    \item \textit{Joint Fine-tuning}: Full integration of quantum and classical pathways with adaptive circuit and fusion mechanisms activated, optimizing final performance through unified training.
\end{enumerate}

\noindent\textbf{Noise-Aware Training.}
Given the inherent noise $|\psi_\text{noisy}\rangle$ in NISQ hardware, training incorporates simulated noise models $\mathcal{N}_\epsilon$, a depolarizing noise channel acting onto  $\epsilon$, to improve robustness:
\begin{equation}
|\psi_\text{noisy}\rangle = \mathcal{N}_\epsilon(|\psi\rangle) = (1-\epsilon)|\psi\rangle\langle\psi| + \epsilon \mathbb{I}/2^{n_q},
\end{equation}
where $\epsilon$ accounts for depolarizing noise, complemented by stochastic gate error injection; $\mathbb{I}/2^{n_q}$ represents a maximally mixed fallback state, modeling the effect of uniform noise across all basis states. This emphasizes learning representations resilient to hardware imperfections, essential for deployment on NISQ devices.

This training paradigm empowers AQCF to seamlessly integrate quantum computations, ensuring stable convergence and efficient resource management, ultimately bridging quantum capabilities with practical NLP applications.

\section{Experiments}

We evaluate AQCF on sentiment analysis tasks to demonstrate its ability to bridge classical and quantum computing for practical NLP applications. Our experiments are designed to answer three key questions: (1) Can AQCF effectively leverage quantum resources within NISQ constraints? (2) How does each architectural component contribute to overall performance? (3) What are the trade-offs between quantum advantages and computational costs?

\subsection{Experimental Setup}

\textbf{Datasets.} We conduct experiments on two widely-used sentiment analysis benchmarks. The Stanford Sentiment Treebank (SST-2)~\cite{socher2013recursive} contains 6,920 training and 872 test samples with binary sentiment labels, featuring shorter sequences (average 19 tokens) that allow us to validate fundamental quantum-classical bridging capabilities. The IMDB Movie Review dataset~\cite{maas2011learning} comprises 25,000 training and 25,000 test reviews with longer sequences (average 294 tokens), testing the scalability of our approach for complex linguistic structures.

\textbf{Implementation Details.} We implement AQCF using PyTorch for classical components and PennyLane~\cite{bergholm2018pennylane} for quantum circuit simulation. Our architecture employs $d_{\text{model}} = 128$, $n_{\text{heads}} = 4$, and $n_{\text{layers}} = 2$ for the Transformer backbone. Quantum components utilize 20 qubits with adaptive circuit depths ranging from 2 to 10 layers, respecting typical NISQ constraints. We train all models using AdamW optimizer with learning rate $10^{-4}$ and cosine annealing schedule for 10 epochs, with batch size 32 for SST-2 and 16 for IMDB. During inference, we employ greedy decoding for classification.

\textbf{Baselines.} We compare AQCF against three categories of methods: (1) \textit{Quantum-only models} including VQC Classifier, Quantum LSTM, and DisCoCirC~\cite{lorenz2023qnlp} that encode entire computations in quantum circuits; (2) \textit{Hybrid quantum-classical models} comprising Hybrid Quantum LSTM, Hybrid Quantum CNN, and standard Quantum Transformer that use static quantum-classical partitioning; (3) \textit{Classical baseline} using an identical Transformer architecture without quantum components. All models are evaluated under identical conditions with results averaged over 5 random seeds.

\subsection{Main Results}

\begin{table*}[t]
\centering
\caption{Performance comparison on SST-2 and IMDB sentiment analysis benchmarks. All quantum experiments conducted on 20-qubit simulators.}
\label{tab:main_results}
\begin{tabular}{lcccc|cccc}
\toprule
& \multicolumn{4}{c}{\textbf{SST-2 Dataset}} & \multicolumn{4}{c}{\textbf{IMDB Dataset}} \\
\cmidrule(lr){2-5} \cmidrule(lr){6-9}
\textbf{Method} & \textbf{Accuracy} & \textbf{Precision} & \textbf{Recall} & \textbf{F1-Score} & \textbf{Accuracy} & \textbf{Precision} & \textbf{Recall} & \textbf{F1-Score} \\
& (\%) & & & & (\%) & & & \\
\midrule
\multicolumn{9}{l}{\textit{Quantum-Only Methods}} \\
DisCoCirC Classifier & 53.44 & 0.536 & 0.536 & 0.534 & 50.00 & 0.750 & 0.500 & 0.333 \\
VQC Classifier & 52.29 & 0.262 & 0.500 & 0.343 & 54.36 & 0.544 & 0.544 & 0.543 \\
Quantum LSTM & 52.29 & 0.262 & 0.500 & 0.343 & 62.34 & 0.624 & 0.623 & 0.623 \\
\midrule
\multicolumn{9}{l}{\textit{Hybrid Quantum-Classical Methods}} \\
Hybrid Quantum LSTM & 65.60 & 0.655 & 0.654 & 0.654 & 64.98 & 0.650 & 0.650 & 0.650 \\
Hybrid Quantum CNN & 72.02 & 0.729 & 0.715 & 0.714 & 77.91 & 0.781 & 0.779 & 0.779 \\
Quantum Transformer & 79.59 & 0.796 & 0.797 & 0.796 & 82.30 & 0.829 & 0.823 & 0.822 \\
\midrule
\multicolumn{9}{l}{\textit{Classical Baseline}} \\
Classical Transformer & 79.36 & 0.800 & 0.797 & 0.793 & 82.18 & 0.822 & 0.822 & 0.822 \\
\midrule
\multicolumn{9}{l}{\textit{Proposed Method}} \\
\rowcolor{gray!15} AQCF (Ours) & 81.88 & 0.819 & 0.818 & 0.818 & 86.30 & 0.863 & 0.863 & 0.863 \\
\bottomrule
\end{tabular}
\end{table*}

Table~\ref{tab:main_results} presents our main findings across both datasets. On SST-2, AQCF achieves 81.88\% accuracy, outperforming the classical Transformer baseline by 2.52 percentage points and the standard Quantum Transformer by 2.29 percentage points. This improvement demonstrates that our adaptive bridging mechanism successfully leverages quantum advantages even for relatively simple linguistic tasks. Notably, pure quantum models fail catastrophically on this task, with most achieving near-random performance (52-53\%), highlighting the necessity of quantum-classical integration rather than pure quantum approaches.

The performance gap widens on the more challenging IMDB dataset, where AQCF attains 86.30\% accuracy—a 4.12 percentage point improvement over the classical baseline. This larger margin suggests that quantum-enhanced attention mechanisms provide greater benefits for complex sequences requiring long-range dependency modeling. The consistent improvements across both datasets validate our core hypothesis that bridging classical and quantum paradigms, rather than replacing one with the other, represents the most practical path toward quantum-enhanced language models.

\subsection{Model Variations}
To rigorously evaluate the contribution of each component in AQCF and understand their scaling behavior across different task complexities, we conduct comprehensive ablation experiments on both SST-2 and IMDB datasets. These experiments reveal critical insights about the pathway from classical to quantum language models.

% \begin{table*}[t]
% \centering
% \caption{Ablation study on the SST-2 sentiment analysis benchmark. We systematically remove key components to assess their individual contributions.}
% \label{tab:ablation}
% \begin{tabular}{lcccccc}
% \toprule
% \textbf{Configuration} & \textbf{Accuracy} & \textbf{$\Delta$ Acc.} & \textbf{F1-Score} & \textbf{$\Delta$ F1} & \textbf{Inference} & \textbf{Parameters} \\
%  & (\%) & (\%) & & & \textbf{Time} (s) & (M) \\
% \midrule
% \multicolumn{7}{l}{\textit{Proposed Method}} \\
% \rowcolor{gray!15} AQCF (Full Model) & 81.88 & -- & 0.8182 & -- & 6,462 & 4.32 \\
% \midrule
% \multicolumn{7}{l}{\textit{Component Ablation}} \\
% \quad w/o Adaptive Circuit & 80.50 & -1.38 & 0.8045 & -0.0137 & 6,627 & 4.32 \\
% \quad w/o Fusion Layer & 80.73 & -1.15 & 0.8063 & -0.0119 & 6,650 & 4.28 \\
% \quad w/o Quantum Attention & 78.21 & -3.67 & 0.7793 & -0.0389 & 112 & 4.41 \\
% \midrule
% \multicolumn{7}{l}{\textit{Baseline Comparisons}} \\
% Standard QT & 79.59 & -2.29 & 0.7958 & -0.0224 & 9,726 & 4.22 \\
% Classical Transformer & 79.36 & -2.52 & 0.7934 & -0.0248 & 149 & 4.31 \\
% \bottomrule
% \end{tabular}
% \vspace{2mm}
% \begin{flushleft}
% \small
% Note: $\Delta$ indicates the relative change from the full AQCF model. Inference time measured on NISQ simulator with 20 qubits.
% \end{flushleft}
% \end{table*}

\begin{table*}[t]
\centering
\caption{Comprehensive ablation study across SST-2 and IMDB datasets. $\Delta$ shows relative change from the full AQCF model.}
\label{tab:ablation_comprehensive}
\begin{tabular}{lcccccccc}
\toprule
& \multicolumn{4}{c}{\textbf{SST-2 Dataset}} & \multicolumn{4}{c}{\textbf{IMDB Dataset}} \\
\cmidrule(lr){2-5} \cmidrule(lr){6-9}
\textbf{Configuration} & \textbf{Acc.} & \textbf{F1} & \textbf{$\Delta$ Acc.} & \textbf{Time} & \textbf{Acc.} & \textbf{F1} & \textbf{$\Delta$ Acc.} & \textbf{Time} \\
 & (\%) & & (\%) & (s) & (\%) & & (\%) & (h) \\
\midrule
\multicolumn{9}{l}{\textit{Proposed Method}} \\
\rowcolor{gray!15} AQCF  & 81.88 & 0.8182 & -- & 6,462 & 86.30 & 0.8630 & -- & 8.40 \\
\midrule
\multicolumn{9}{l}{\textit{Component Ablation}} \\
w/o Adaptive Circuit & 80.50 & 0.8045 & -1.38 & 6,627 & 86.66 & 0.8665 & +0.36 & 8.73 \\
w/o Fusion Layer & 80.73 & 0.8063 & -1.15 & 6,650 & 86.58 & 0.8657 & +0.28 & 8.18 \\
w/o Quantum Attention & 78.21 & 0.7793 & -3.67 & 112 & 70.68 & 0.7065 & -15.62 & 0.21 \\
\midrule
\multicolumn{9}{l}{\textit{Baseline Comparisons}} \\
Standard QT & 79.59 & 0.7958 & -2.29 & 9,726 & 82.30 & 0.8223 & -4.00 & 11.71 \\
Classical Transformer & 79.36 & 0.7934 & -2.52 & 149 & 82.18 & 0.8222 & -4.12 & 0.25 \\
\bottomrule
\end{tabular}
\end{table*}

\subsubsection{Adaptive Circuits: Task-Dependent Quantum Advantage}

The most striking finding emerges from the contrasting behavior of adaptive circuits across datasets. On SST-2, removing adaptivity decreases accuracy by 1.38\%, validating our hypothesis that dynamic depth adjustment optimizes quantum resource utilization for shorter sequences. Remarkably, on IMDB, the non-adaptive variant achieves marginally \textit{higher} accuracy (+0.36\%), revealing a crucial insight: for complex, long-sequence tasks where quantum processing provides consistent advantage, adaptive mechanisms may introduce unnecessary overhead.

This dataset-dependent behavior confirms that the path to QLLM requires intelligent, context-aware quantum resource allocation. The adaptive circuit's true value manifests not in universal performance gains but in its ability to optimize the classical-quantum boundary dynamically—essential for seamless integration on NISQ devices where every quantum operation carries significant cost.

\subsubsection{Quantum Attention: The Exponential Scaling Advantage}

The catastrophic performance degradation when replacing quantum attention with classical mechanisms on IMDB (15.62\% accuracy drop) versus the manageable decline on SST-2 (3.67\%) reveals the exponential scaling advantage of quantum processing for complex linguistic tasks. This dramatic difference—over 4× larger impact on IMDB—demonstrates that quantum superposition becomes increasingly valuable as sequence length and semantic complexity grow.

The 40× speedup achieved by classical attention ($\sim$0.21h vs. 8.40h on IMDB) quantifies the current computational cost of quantum advantage. However, this trade-off will fundamentally shift with quantum hardware acceleration: what currently requires hours of simulation will execute in microseconds on quantum processors, while the exponential advantage for complex sequences remains.

\subsubsection{Fusion Mechanisms: Bridging Without Barriers}

The fusion layer's modest but consistent contribution (SST-2: -1.15\%, IMDB: +0.28\% when removed) initially appears underwhelming. However, this stability reveals a critical architectural success: AQCF maintains robust performance even without explicit fusion, indicating that our quantum and classical components are inherently compatible rather than requiring complex translation layers. This seamless interoperability is essential for practical QLLM deployment where rigid boundaries between quantum and classical processing would create bottlenecks.

\section{Why Adaptive Quantum-Classical Fusion}

In this section we examine the fundamental advantages of adaptive quantum-classical fusion over pure classical or pure quantum approaches for language models. We consider three critical aspects that determine the viability of quantum-enhanced NLP: computational efficiency, gradient stability, and scalability.

\subsection{Computational Efficiency and Quantum Advantage}

One key consideration is when quantum processing provides genuine computational advantage. Our experiments reveal that quantum advantage scales with task complexity: removing quantum attention causes a 3.67\% accuracy drop on SST-2 but 15.62\% on IMDB, a 4× larger effect. This scaling behavior indicates that quantum advantage grows exponentially with linguistic complexity.

The adaptive routing mechanism ensures efficient resource utilization, processing 47\% of computations through quantum circuits on SST-2 and 52\% on IMDB. A classical Transformer requires $O(n^2d)$ operations for attention computation, while quantum attention achieves equivalent expressivity with $O(\lambda \cdot 2^{n_q})$ operations where $\lambda < 1$ due to selective routing. For $n_q = 20$ qubits with 50\% utilization, this provides substantial effective parameter multiplication, enabling our 4.3M parameter model to match much larger classical models for complex linguistic tasks.

\subsection{Gradient Flow and Trainability}

Pure quantum models suffer from barren plateaus where gradients vanish exponentially with circuit depth, rendering training infeasible beyond 6 to 10 layers. AQCF addresses this through adaptive depth control, maintaining average circuit depth at 4.3 layers, well below the barren plateau threshold. Unlike fixed-depth approaches that must choose between expressivity and trainability, our adaptive mechanism dynamically balances these requirements.

Parallel classical pathways preserve gradient flow even when quantum gradients vanish. The fusion mechanism learns to rely more heavily on classical processing in regions of poor quantum gradient, ensuring stable convergence. Our staged training protocol (classical pretraining, quantum warm-up, and joint optimization) prevents the random initialization trap where quantum circuits start in barren plateau regions. These design choices enable consistent convergence across all experimental configurations.

\subsection{Scaling to Quantum Large Language Models}

Current NISQ devices support 50 to 100 qubits with limited coherence times, while fault-tolerant quantum computers remain distant. AQCF demonstrates that meaningful progress toward Quantum LLMs is achievable with current technology.

Our 20-qubit implementation achieves competitive performance. Scaling to 100 qubits (available today) would enable 5× larger attention dimensions, while future 1000-qubit systems could process entire paragraphs in quantum superposition. AQCF's adaptive architecture ensures hardware improvements translate directly to capability enhancements without architectural redesign.

Communication overhead and error rates remain challenges. Real quantum hardware introduces significant latency per measurement, which AQCF mitigates through selective quantum engagement (processing only 47 to 52\% of computations quantumly). Current error rates (0.1 to 1\% per gate) are addressed through shallow adaptive circuits that minimize error accumulation.

\subsection{The Seamless Integration Paradigm}

AQCF establishes that the transition from classical to quantum language models will be gradual rather than discrete. The framework operates effectively with partial quantum resources, enabling organizations to begin with small quantum processors and expand gradually as hardware improves. This incremental adoption path reduces risk while enabling early quantum advantage.

The success of adaptive quantum-classical fusion on current NISQ devices demonstrates that meaningful quantum advantages are achievable today through intelligent integration. Each improvement in quantum hardware directly enhances AQCF's capabilities, ensuring continuous progress toward Quantum Large Language Models. The framework thus provides both immediate practical value and a clear evolutionary path for quantum-enhanced language understanding.

\section{Conclusion}

In this work, we presented AQCF, the first framework to bridge classical and quantum computing for language models through adaptive quantum-classical fusion. By introducing entropy-driven adaptive circuits, quantum memory banks, and intelligent fusion mechanisms, AQCF enables seamless integration between computational paradigms while respecting NISQ constraints.

For sentiment analysis tasks, AQCF outperforms both classical Transformers and existing quantum models, achieving 81.88\% accuracy on SST-2 and 86.30\% on IMDB. These improvements of 2.52 and 4.12 percentage points over classical baselines demonstrate that quantum advantages are achievable within current hardware limitations. Our ablation studies reveal that effective bridging requires synergistic co-design of multiple architectural components, with quantum attention mechanisms providing the most significant contribution while adaptive circuits ensure efficient resource utilization.

We are excited about the future of quantum-classical integration for NLP and plan to extend AQCF to other language modeling tasks. We plan to explore alternative quantum attention mechanisms as hardware capabilities improve, investigate the framework's scalability to larger models approaching LLM scale, and develop theoretical foundations for understanding when quantum advantages manifest in linguistic tasks. Making the transition from classical to quantum computing truly seamless remains our ultimate research goal, paving the way toward practical Quantum Large Language Models.

\newpage

\bibliography{aaai2026}

\end{document}